\documentclass[twocolumn,aps,showpacs,preprintnumbers,amsmath,amssymb,floatfix,superscriptaddress,pra]{revtex4-1}
\usepackage{graphicx}
\usepackage{amsmath,amsfonts,amssymb}
\usepackage{graphicx}
\usepackage{color}
\usepackage{subcaption}

\def\beq{\begin{equation}}
\def\eeq{\end{equation}}

\newcommand{\affW}{Faculty of Physics, University of Warsaw, Pasterua 5, 02-093 Warsaw, Poland}
\newcommand{\affD}{Zentrum f\"ur Optische Quantentechnologien and The Hamburg Centre for Ultrafast Imaging, Universit\"at Hamburg, Luruper Chaussee 149, D-22761 Hamburg, Germany}
\newcommand{\affE}{Forschungszentrum J\"{u}lich Gmbh, Institute of Quantum Control (PGI-8), D-52425 J\"{u}lich, Germany}

\begin{document}

\title{
Quantum simulation of extended polaron models using compound atom-ion systems
}
\author{Krzysztof~Jachymski}\affiliation{\affE,\affW}
\author{Antonio~Negretti}\affiliation{\affD}

\begin{abstract}
We consider the prospects for quantum simulation of condensed matter models exhibiting strong electron-phonon coupling using a hybrid platform of trapped laser-cooled ions interacting with an ultracold atomic gas. This system naturally possesses a phonon structure, in contrast to the standard optical lattice scenarios usually employed with ultracold atoms in which the lattice is generated by laser light and thus it remains static. We derive the effective Hamiltonian describing the general system and discuss the arising energy scales, relating the results to commonly employed extended Hubbard-Holstein models. Although for a typical experimentally realistic system the coupling to phonons turns out to be small, we provide the means to enhance its role and reach interesting regimes with competing orders. Extended Lang-Firsov transformation reveals the emergence of phonon-induced long-range interactions between the atoms, which can give rise to both localized and extended bipolaron states with low effective mass, indicating the possibility of fermion pairing.

\end{abstract}

\date{\today}

\maketitle

%
\section{Introduction.} 
%
Strong interactions in quantum many-body systems can lead to exotic collective effects that are difficult to characterize and understand at the microscopic level. The combination of the complexity arising from the Hilbert space exponentially growing with the particle number and the inherent entanglement of the many-body wave functions limit the power of even state-of-the-art computational methods. For these reasons, quantum simulation~\cite{Cirac2012} has emerged as an alternative approach, aiming to create highly controllable artificial systems that would be well understood microscopically and easy to scale in the number of qubits. Multiple physical platforms have been developed with this task in mind, including superconducting circuits, photonic systems, trapped ions and ultracold atoms~\cite{Bloch2012,Blatt2012,Aspuru2012,Houck2012}.

One major challenge in condensed matter physics is connected with the interplay of strong electron correlations and large, possibly finite-range, electron-phonon coupling~\cite{lanzara2001evidence}. Such systems have been theoretically studied for a long time~\cite{Frohlich1954,Linden2004,Hohenadler2007,Devreese2009} and are believed to play an essential role in the formation of the superconducting state in certain materials~\cite{Alexandrov1981,Alexandrov1986,Alexandrov2011}. The physical picture behind the phenomenon can be provided by introducing polarons, which are quasiparticles composed of electrons dressed with lattice phonons. Their mutual interaction can lead to the formation of bound states with low effective mass that would thus be mobile and Bose condense at high temperature, leading possibly to high-$T_c$ superconductivity~\cite{Scalapino2018}. However, this might require carefully tuned system parameters, and in general polaron models can feature rich physics depending on the system geometry and the type of interactions~\cite{Alexandrov2010}. In the strong coupling limit, analytical predictions for the polaron and bipolaron properties such as their effective mass can be derived~\cite{Bonvca1999,Bonvca2001}. While the research on the static properties of different polaron models is still active and fruitful~\cite{Swartz2018,Sous2018,Reinhard2019,Wang2019}, there is a growing interest in polaron dynamics out-of-equilibrium. A particularly interesting scenario to consider in this context is the light-induced superconducting response of the system, which
has been observed in several materials~\cite{Fausti2011,Subedi2014,Mitrano2016,Babadi2017,Murakami2017,Wang2018a,Cavalleri2018}. The understanding of high-$T_c$ superconductivity remains incomplete, in particular on the microscopic level. Nonetheless, enormous advances in providing insight into the system properties via photonic spectroscopy of correlated materials in- and out-of-equilibrium have been attained both theoretically and experimentally~\cite{Wang2018}. 

Quantum simulations of polaron physics have been proposed theoretically using ultracold atomic mixtures~\cite{Jorgensen2016,Hu2016,Cetina2016,Scazza2017,Camargo2018}, Rydberg atoms~\cite{Hague2012a,Hague2012b}, cold molecules~\cite{Pupillo2008,Ortner2009,Herrera2013}, trapped ions~\cite{Stojanovic2012} and atom-ion systems~\cite{CasteelsJLTP11,Gregory2020}. Particularly the latter opens intriguing perspectives, as ions confined in radiofrequency traps form crystal structures that combined with fermionic atoms emulate a solid-state material naturally, contrarily to atoms trapped in optical 
lattices~\cite{Jaksch1998,Bloch2005}, for which the back-action of the atoms on the lattice potential is very weak. Given  the exceptional control of preparation and measurement of trapped ion systems~\cite{Leibfried2003}, especially of their motion, the compound atom-ion system represents a promising candidate for quantum simulation of solid-state physics~\cite{Bissbort2013}, including extended Hubbard models~\cite{Negretti2014}, lattice gauge theories~\cite{Dehkharghani2017}, topological states~\cite{Gonzalez2018}, and charge transport~\cite{Cote2000,Mukherjee2019,dieterle2020transport} (see Ref.~\cite{TomzaRMP} for a detailed review). On the experimental side, various strategies have been developed in order to reach the quantum regime with these systems. In the most standard state-of-the-art setting utilizing radiofrequency ion traps, a small atom-to-ion mass ratio (e.g., lithium atoms and ytterbium ions) enables one to reach the $s$-wave collision energy~\cite{Feldker2019}. Successful sympathetic cooling of the ion has also been reported in optical traps~\cite{Karpa2020}. Furthermore, it has been shown that sub-microkelvin temperatures can be attained when ionizing a Rydberg atom inside a Bose-Einstein condensate~\cite{Kleinbach2018}. These pioneering experiments open the door to study the aforementioned electron-phonon physics in the near future.

In this paper, we investigate the formation of bipolarons in extended Hubbard-Holstein models (HHM) that can be engineered with atom-ion systems. Specifically, we consider a linear ion crystal superimposed with a degenerate Fermi gas. Even in such a one-dimensional setting interesting features are expected. For instance, it has been predicted that at half filling a metallic phase in the HHM emerges as a result of the competition of the charge- and spin-density wave orders~\cite{Takada2003,Clay2005,Tezuka2007,Fehske2008,Campbell2014}. Furthermore, in the strong electron-phonon coupling regime, a long-ranged (i.e., nonlocal) electron-phonon coupling decreases the effective mass of polarons and bipolarons, thus enhancing the mobility of those quasi-particles~\cite{Bonvca2001,Sous2018}. We show that the atom-phonon coupling in a compound atom-ion system can be made tunable with experimentally realistic techniques. Compared to previous studies~\cite{Bissbort2013,Negretti2014,Michelsen2019}, here we treat the ions quantum mechanically and show how the resulting atom-phonon coupling can be exploited to form bipolarons. Our findings pave the way towards quantum simulation of extended HHM with tunable long-ranged atom-phonon couplings, and therefore realization of interesting quantum phases in the laboratory. 

%
\section{System and effective Hamiltonian}
%

\begin{figure*}
	\centering
	\begin{subfigure}[b]{0.3\textwidth}
		\includegraphics[width=\textwidth]{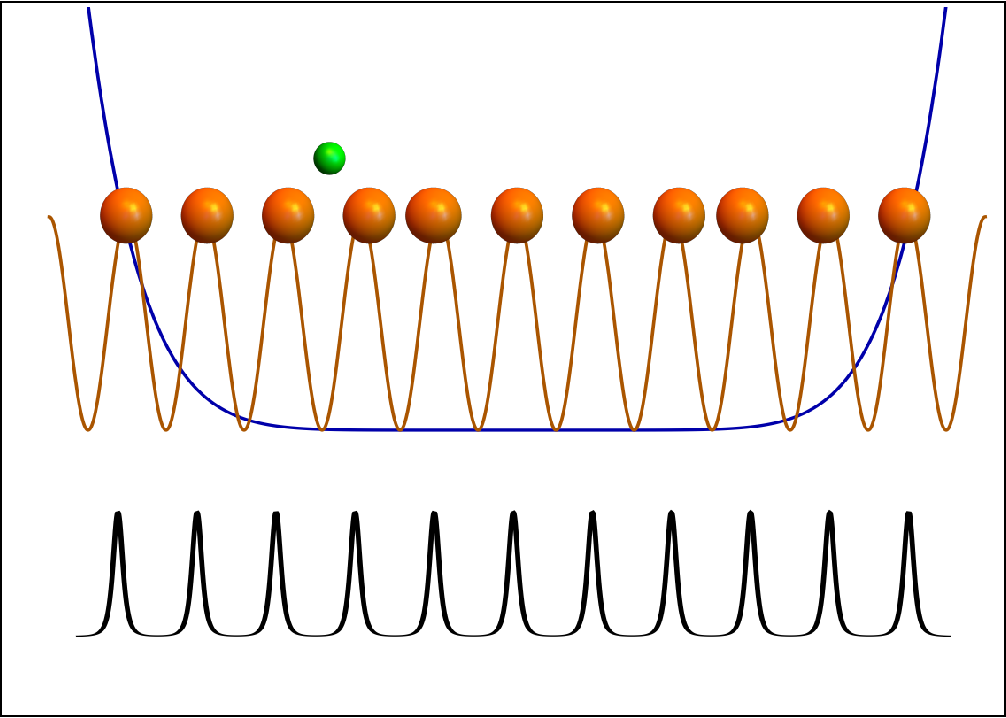}
	\end{subfigure}
	\begin{subfigure}[b]{0.24\textwidth}
		\includegraphics[width=\textwidth]{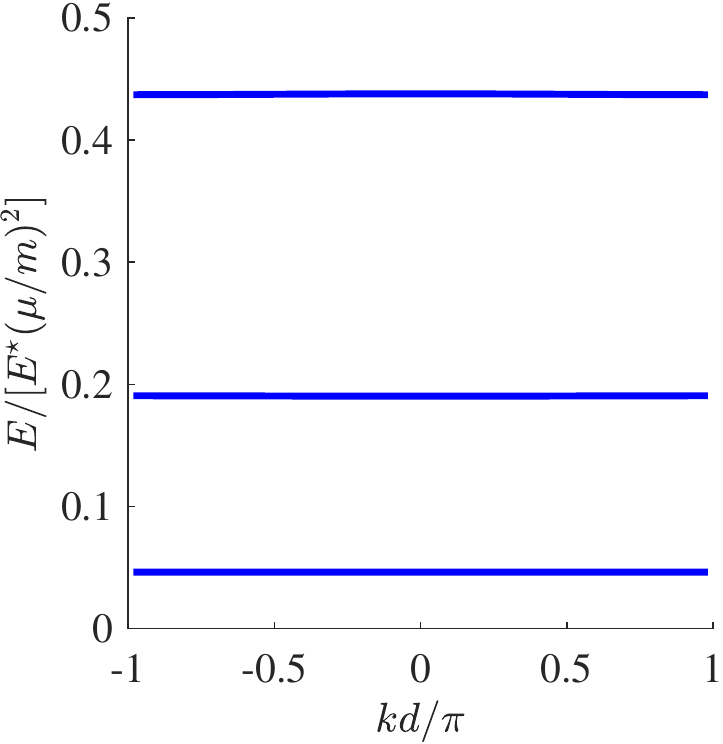}
	\end{subfigure} 
	\begin{subfigure}[b]{0.35\textwidth}
		\includegraphics[width=\textwidth]{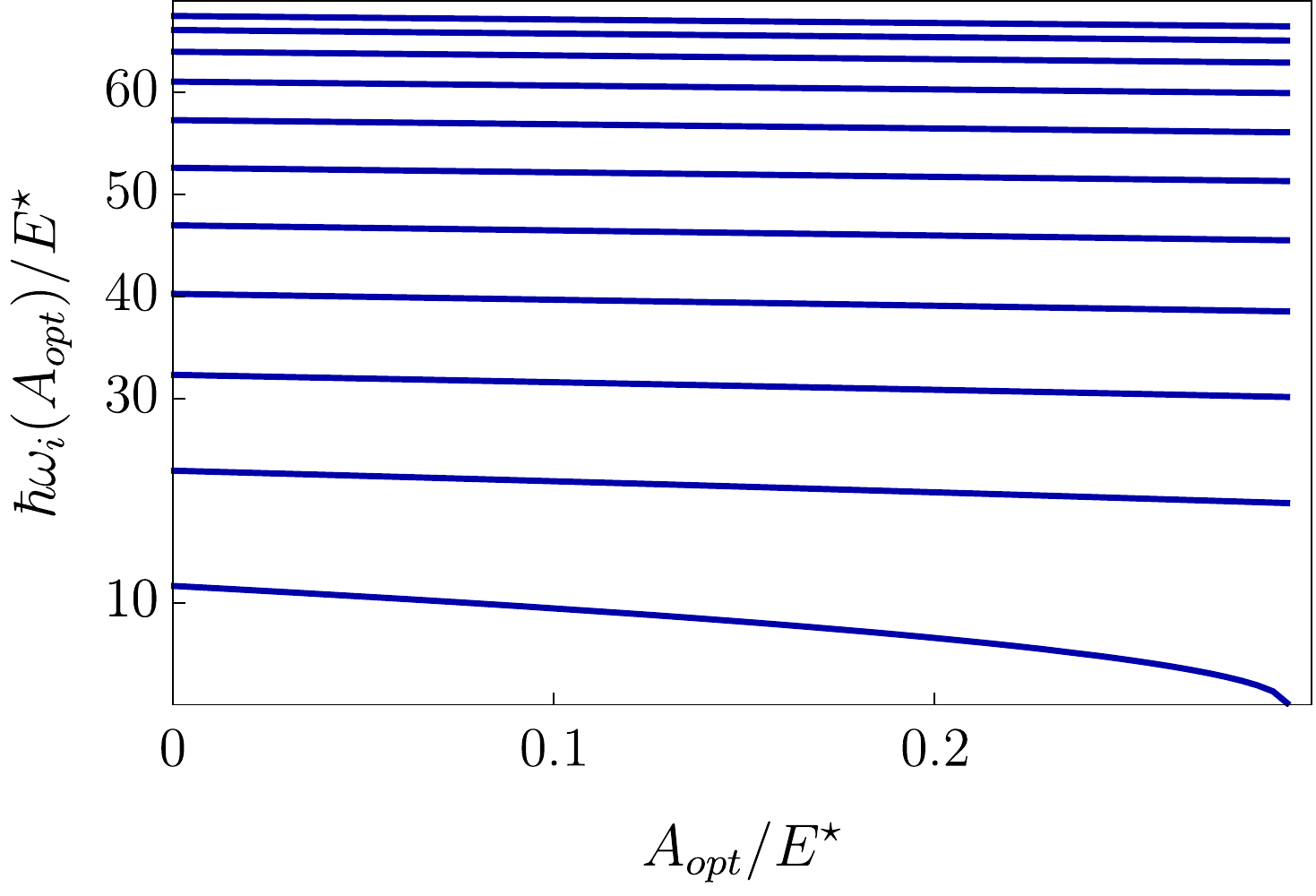}
	\end{subfigure}
	\caption{\label{fig:setup}(a) Schematic drawing of the system consisting of an ion chain (big orange balls) in an external trap (blue line) and a repulsive optical lattice potential (orange line). A neutral atom (small green ball) is moving in a periodic potential stemming from the interaction with the ion chain (black line below). (b) Typical band structure of a single atom moving in the ionic lattice with strong ion-atom interactions. (c) Phonon spectrum of a finite linear ion chain consisting of $N=11$ ions as a function of the external lattice depth $A_{\rm opt}$. In panels (b) and (c) the ion separation is $d = 15 R^\star$.}
\end{figure*}

We consider an ensemble of identical ions in an external trap and a gas of fermionic atoms overlapping with it, schematically presented in Fig.~\ref{fig:setup}(a). To characterize the ionic part, we first minimize the classical energy functional consisting of the  trapping potential and Coulomb interaction with respect to the ions' positions. In general, the interplay between the trap and the interactions can lead to various system geometries with structural phase transitions between them. Here, we assume that the ion chain is linear and thermodynamically stable so that small displacements of the ions from the equilibrium only give rise to phonon excitations. A convenient approach to calculate the phonon spectrum in the general case has been provided, e.g., in Ref.~\cite{Bissbort2016} (see Appendix~\ref{appa} for more details). Each ion can be associated with its local harmonic oscillator frequency defined as $\Omega_j=\sqrt{\frac{V_{jj}}{M}}$, where $V_{ij}=\frac{\partial^2 V}{\partial \left(\delta R_i\right)\partial \left(\delta R_j\right)}$ is the second derivative of the total potential energy calculated at equilibrium and $M$ denotes the mass of the ion, while $\delta R_j$ is the displacement of the $j$-th ion from its equilibrium position. In the next step one introduces local ladder operators corresponding to these local oscillators and rewrites the Hamiltonian. The latter acquires quadratic form and can be diagonalized using a generalized Bogoliubov transformation, leading to the phonon mode structure
\beq
\hat H_{\rm ion}=\sum_m{\hbar\omega_m \hat b_m^\dagger \hat b_m}
\eeq
with $\omega_m$ being the energy of the $m$-th collective mode, and $\hat b_m$, $\hat b_m^\dagger$ denoting the phonon creation and annihilation operators that fulfill the usual bosonic commutation relations. For a finite ion number the spectrum is discrete and thus gapped, as shown in Fig.~\ref{fig:setup}(c), while for an infinitely long chain it becomes continuous with an acoustic branch along the chain and an optical one in the transverse directions~\cite{Fishman2008}. Crucially, the ions can be individually addressed and driven using additional optical pulses~\cite{Blatt2012}, allowing for some degree of manipulation of the phonon structure and dynamics, including creation of squeezed states~\cite{Kienzler2017,FossFeig2019}. 

The gas of neutral atoms of mass $m$ is assumed to be confined in a quasi-one-dimensional waveguide, with transverse confinement being sufficiently strong to freeze the atomic motion in the ground state. Albeit the gas can be in principle of either bosonic or fermionic nature, in this work we focus on the fermionic case only.
The atom-ion interaction at large distances from the ion core (typically above a few nanometers) is given by the polarization potential 
\begin{align}
\label{eq:Vai}
V_{\rm ai}(r) = - \frac{C_4}{r^4}
\end{align}
with $r=\vert \mathbf{r}\vert$ being the separation between the atom and the ion and $C_{4} = \alpha e^2/(8\pi\epsilon_0)$ -- in SI units -- with $\alpha$ being the static atom polarizability, $e$ the electron charge, and $\epsilon_0$ the vacuum permittivity. The potential is characterized by the length $R^\star$ and energy $E^\star$ scales
\begin{align}
R^\star = \sqrt{\frac{2\mu C_4}{\hbar^2}}, \qquad 
E^\star = \frac{\hbar^2}{2 \mu (R^\star)^2},
\label{eq:units}
\end{align}
where $\mu$ is the reduced mass $\mu=Mm/(m+M)$. A possible choice for the atom-ion pair is $^6\text{Li}$ / $^{174}\text{Yb}^+$, which due to the low mass ratio is the most favorable to attain the ultracold regime in radio-frequency traps~\cite{Cetina2012,Joger2014,Feldker2019}. For this pair we have $E^\star/h \simeq 178.6 \text{ kHz}$, $R^\star \simeq 69.8 \text{ nm}$, and the mass ratio of about 0.035. In the lowest order, one can assume the ions to be static and the atoms thus move in a periodic potential resulting from the interaction with the ions depicted as a black line in Fig.~\ref{fig:setup}(a). As long as the spacing between the ions is much bigger than $R^\star$, which is a reasonable assumption as in a typical experimental setup the ion spacing reaches a few $\mu$m, it is sufficient to use the one-dimensional pseudopotential approximation for $V_{\rm a-i}$~\cite{Negretti2014}
\beq
\label{eq:pseudo}
V_{\rm a-i}(x)=g^e\delta(x)+g^o\delta^\prime(x)\partial_\pm
\eeq
with coefficients $g^e$, $g^o$ describing the interaction in the even and odd partial waves (one-dimensional analogues of the three-dimensional case). 
The action of the operator on the right of Eq.~(\ref{eq:pseudo}) on a test function is defined as: $2\, \hat\partial_{\pm}\psi(x) = [\psi^{\prime}(0^+) + \psi^{\prime}(0^-)]$ with $\psi^{\prime}(0^{\pm}) = \lim_{x\rightarrow 0^{\pm}}\psi^{\prime}(x)$, where the apex $^{\prime}$ denotes the spatial derivative.
The periodic potential gives rise to a band structure which we calculate numerically, showing an example in Fig.~\ref{fig:setup}(b). In addition, the atoms are interacting with each other via van der Waals forces which have local character described by a pseudopotential similar to~\eqref{eq:pseudo}. Having calculated the band structure and the corresponding Bloch states, one can switch to the basis consisting of maximally localized Wannier states in which the atomic part of the Hamiltonian takes the familiar form of the extended fermionic Hubbard model~\cite{Negretti2014}
\beq
\hat{H}_a=\sum_{ij\sigma}{J_{ij}\hat{c}_{i\sigma}^\dagger \hat{c}_{j\sigma}}+\sum_{i}{U \hat{n}_{i\uparrow}\hat{n}_{i\downarrow}}+\sum_{ij\sigma\sigma^\prime}{V_{ij} \hat{n}_{i\sigma}\hat{n}_{j\sigma^\prime}}\, .
\eeq
Here the $\hat{c}_{i\sigma}$ are atomic annihilation operators for lattice site $i$, with indices $\sigma ,\,\uparrow ,\,\downarrow$ denoting the two atomic spin states. The next-neighbour terms omitted here are smaller than the leading ones, but typically not completely negligible due to less localized Wannier functions with respect to the case of an optical lattice potential.

%
%

%
\section{Atom-phonon coupling}
\label{sec:atph}
%
The crucial element for the simulation of polaron models is the coupling of the atoms to the phonons which results from the ion-atom interaction beyond the static ion approximation. By expanding to the first order the atom-ion interaction~(\ref{eq:Vai}) with respect to the ions' equilibrium positions, one arrives at the following textbook expression for the atom-phonon coupling 
\beq
V_{\rm a-ph}(\mathbf{r})=\frac{1}{\sqrt{N}}\sum_{k}{V_{ai}(\mathbf{k})e^{i \mathbf{k}\cdot \mathbf{r}}A_k k}.
\label{aphcpl}
\eeq
Here, $N$ is the number of ions, $V_{ai}(\mathbf{k})$ is the lattice Fourier transform of the atom-ion interaction, $\mathbf{k}$ is the lattice quasi-momentum, and $A_k$ is defined as
\beq
A_k \vec{\boldsymbol{\epsilon}}_k=\frac{1}{\sqrt{N}}\sum_j{\delta \mathbf{R}_j e^{-i\mathbf{k R}_j}}\, ,
\eeq
being the lattice Fourier transform of the ion displacement operator with $\mathbf{R}_n$ denoting the equilibrium position of the $n$th ion. 
The atom-phonon coupling term in the Hamiltonian is by definition given as 
\beq
\hat{H}_{\rm a-ph}=\int{d\mathbf{r} \hat \rho(\mathbf{r}) V_{\rm a-ph}(\mathbf{r})}
\eeq
with $\hat{\rho}(\mathbf{r})$ denoting the atomic density operator, which we expand in terms of the lattice Wannier states $w_n$ and perform a Fourier transform, such that
\beq
\hat{\rho}(\mathbf{k})=\sum_{nm}\hat{c}_n^\dagger \hat{c}_m e^{i\mathbf{kR}_n} \alpha_{nm}(\mathbf{k})\, ,
\eeq
where, assuming an effectively one-dimensional system, $\alpha_{nn^\prime}(q)=\int{dy\, w_{n}^\ast(y-R_{nn^\prime})w_n(y) e^{iqy}}$. This results in
\beq
\begin{split}
	\hat H_{\rm a-ph} = -\sqrt{\frac{\hbar}{2MN^2}}\sum_{kjnn^\prime\sigma}{}\frac{\alpha_{nn^\prime}(k) e^{i k R_{nj}}}{\sqrt{\Omega_j}}|k| V_{a-i}(k) \\ \times \hat{c}_{n\sigma}^\dagger \hat{c}_{n^\prime\sigma} \left(\hat{a}_j+\hat{a}_j ^\dagger\right)\, .
\end{split}
\eeq
Here, $R_{nn^\prime}=\left|\mathbf{R}_n-\mathbf{R}_{n^\prime}\right|=d\left|n-n^\prime\right|$, $d$ is the equilibrium distance between the ions and a summation over the phonon modes has been performed to shorten the notation, leading to new local phonon operators connected to a single lattice site (the summation over $j$ cannot be performed in general, since the $\Omega_j$ values may depend on the site index and the system is not translationally invariant)
\beq
\hat{a}_j=\sum_m{(u_j ^m-v_j ^m)\hat{b}_m}\, .
\eeq
The terms leading to tunneling of particles between the lattice sites will in general be suppressed due to the small Wannier function overlap. We can now introduce the local atom-phonon coupling coefficient describing the term that does not involve tunneling (dropping the $n$, $n^\prime$ indices in the $\alpha$ parameter)
\beq
M_{nj}=\sqrt{\frac{\hbar}{2MN\Omega_j}}\frac{1}{\sqrt{N}}\sum_{k}{\alpha(k) e^{ikR_{nj}}|k| V_{e-i}(k)}\, .
\eeq

It is convenient to rewrite this in dimensionless form, using $R^\star, E^\star$ as length and energy units. At this point we also restrict the consideration to the even part of the interaction~\eqref{eq:pseudo}, introducing the even scattering length $a^e=\hbar^2/\mu g^e$, which gives
\beq
\label{eq:mcoeff}
\frac{M_{nj}}{E^\star}=\frac{1}{N}\sqrt{\frac{E^\star}{\hbar\Omega_j}\frac{m}{M}}\sum_k{\alpha(k)\left(|k| R^\star\right)e^{ik R_{nj}}\frac{2m}{\mu}\frac{R^\star}{a^e}\frac{R^\star}{d}}\, .
\eeq
Let us now discuss the interplay of the different quantities present in the model. Expressing the characteristic phonon energy scale $\hbar\bar{\omega}=\sqrt{\frac{\hbar^2 e^2}{4\pi\varepsilon_0 M d^3}}$ in the chosen units leads to 
\beq
\label{eq:phscale}
\frac{\hbar\bar{\omega}}{E^\star}=\sqrt{4\zeta \left(\frac{R^\star}{d}\right)^3 \left(\frac{R^\star}{\xi}\right)\left(\frac{m}{M}\right)}\, ,
\eeq
where $\zeta$ denotes the fine structure constant and $\xi=\hbar/mc$ has the dimension of length. For standard ion-atom pairs and realistic distance $d$ between the ions the number resulting from Eq.~\eqref{eq:phscale} is on the order of 20. At the same time, the tunneling and interaction scales as well as $M_{nj}$ are only a fraction of $E^\star$. This means that typically the phonon dynamics is largely decoupled from the atoms, i.e. the ion chain is stiff. In order to tune the system towards the more interesting regimes, one needs to bring the energy scales closer to each other. The analysis of the $M_{nj}/\hbar\bar{\omega}$ ratio suggests that the most effective solution is to utilize ion-atom Feshbach resonances to increase the $R^\star/a^e$ ratio in Eq.~\eqref{eq:mcoeff}. Further possible control knob is to lower the phonon mode frequencies by shaping the ion trap. One can, for instance, place the ions in an additional optical lattice potential~\cite{CetinaNJP} antialigned with the ions' equilibrium positions $V_{\rm opt}(x)=A_{\rm opt}\cos^2 \left(x/\Lambda\right)$ with the appropriate wavelength $\Lambda$. As long as this does not destabilize the chain, such potential only shifts the mode frequencies to lower energies, which are more compatible with the other terms in the Hamiltonian. The exemplary case showing the mode tuning as a function of the lattice depth $A_{\rm opt}$ for fixed distance between the ions is shown in Fig.~\ref{fig:setup}(c), where we mimick the flat-bottom ion trapping potential realizable e.g. using octupole traps by placing two fixed ions at the edges of the otherwise untrapped 1D chain. We have checked that the lattice has negligible impact on the other parameters, as the ion equilibrium positions are not displaced, so the atomic Wannier functions remain intact. Another reasonable way of tuning the energy scales is to vary the ion separation $d$ using again the external trap structure, as from Eq.~\eqref{eq:mcoeff} one obtains $M_{nj}/\hbar\bar{\omega}\propto d^{5/4}$.

\begin{figure}
	\centering
	\begin{subfigure}[b]{0.45\textwidth}
		\includegraphics[width=0.9\textwidth]{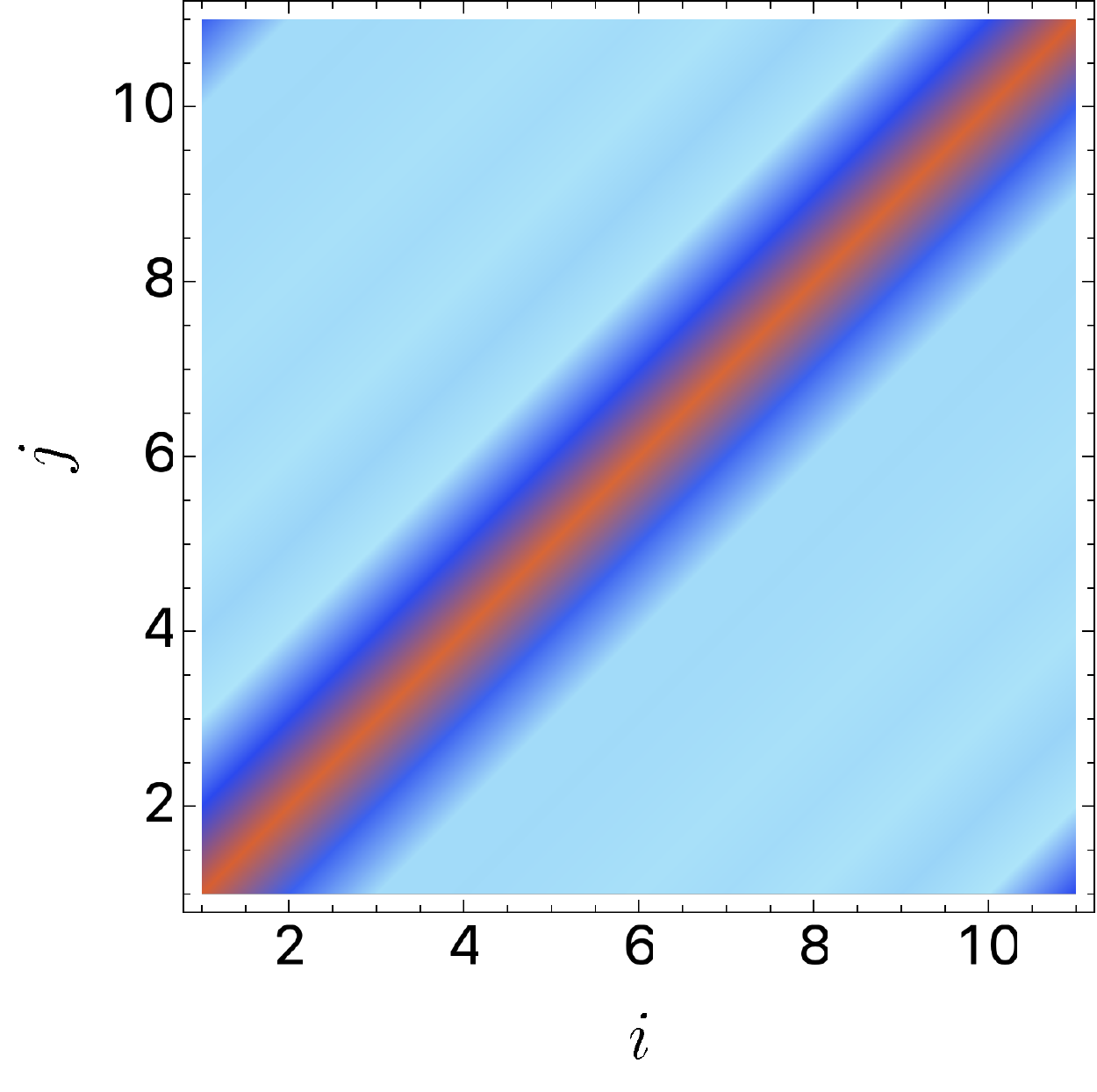}\\
		\includegraphics[width=0.7\textwidth]{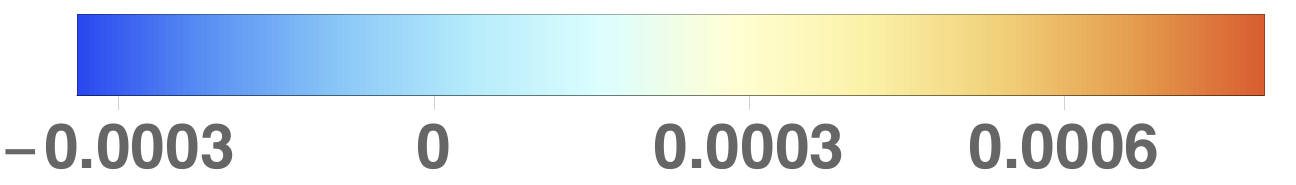}
	\end{subfigure}
\begin{subfigure}[b]{0.45\textwidth}
	\includegraphics[width=0.9\textwidth]{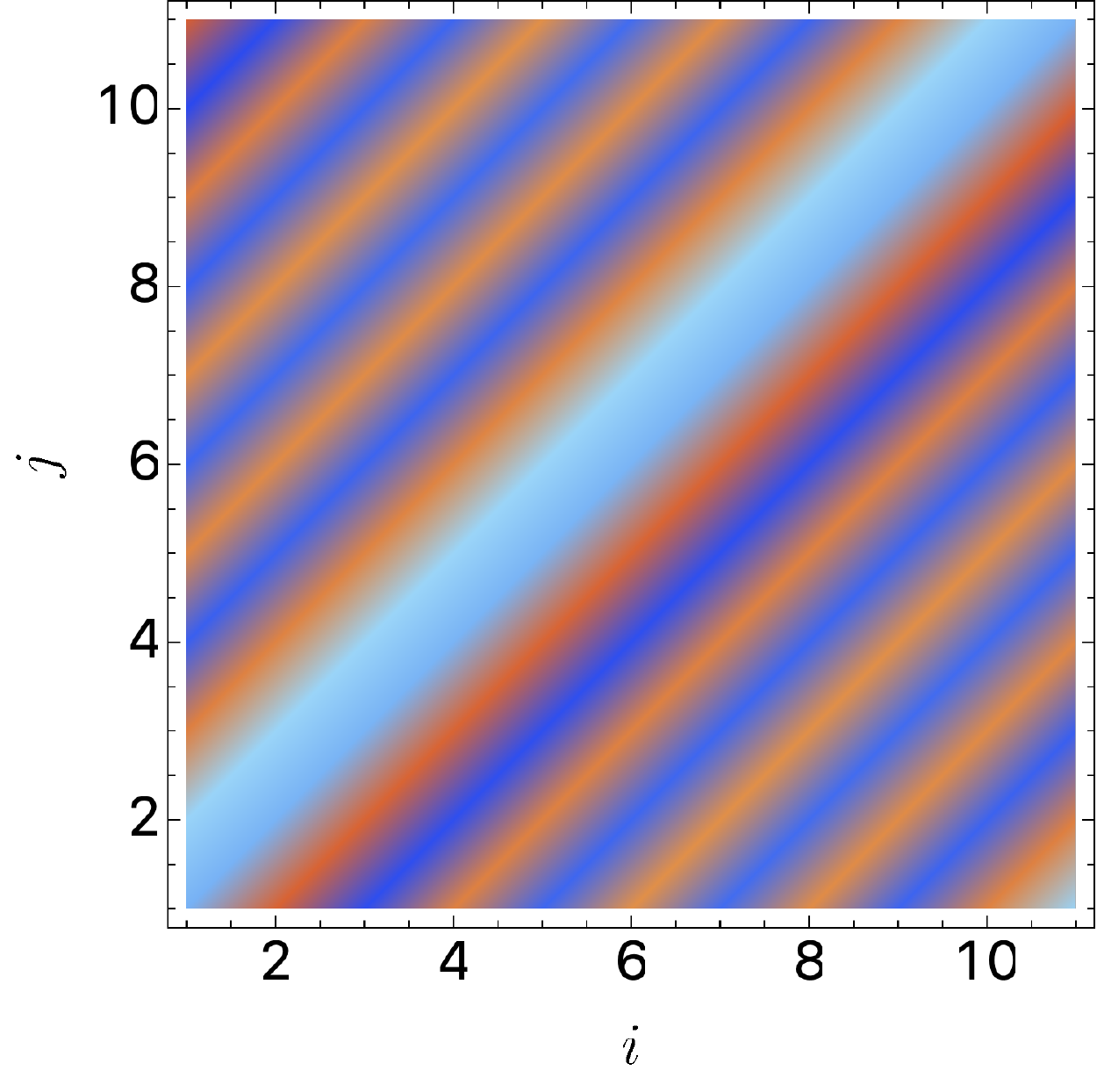}\\
	\includegraphics[width=0.7\textwidth]{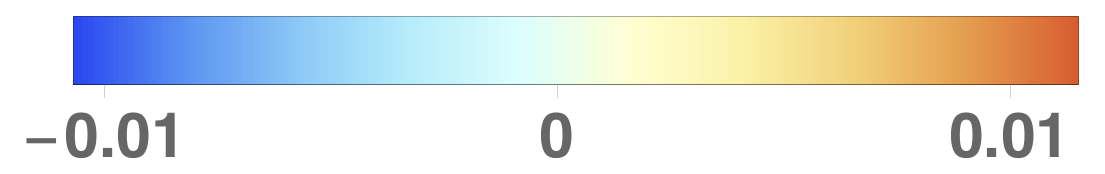}
\end{subfigure}
	\caption{\label{fig:atph}Atom-phonon coupling strength $M_{ij}/E^\star$ in a lattice composed of $N=11$ ions for two different values of the ion-atom scattering length $a^e=0.1R^\star$ (a) and $a^e=0.008R^\star$ (b).}
\end{figure}

Figure~\ref{fig:atph} shows two exemplary cases of the atom-phonon coupling strength $M_{ij}$ in a finite chain consisting of $N=11$ ions. For the presentation we have chosen experimentally realistic parameters corresponding to Yb$^+$ ions and Li atoms with ion separation $d=15\, R^\star$. The difference between the two plots lies in the value of the atom-ion scattering length $a^e$, which for the case depicted in panel (a) takes the value $a^e=0.1\, R^\star$, while for the case (b) $a^e=0.008\, R^\star$, corresponding to strong, resonant interactions. In the latter case we need to include the energy dependence of the scattering length to obtain reliable results. We observe that for weak atom-ion repulsion the resulting coupling is rather local, while for strong interactions it extends over the whole chain, allowing for realization of different regimes.

%
\section{Lang-Firsov transformation}
%
Having discussed the system parameters, we now proceed to the analysis of the connection between the ion-atom simulator and the theoretical models of lattice polarons. As outlined above, the system can be described with the following Hamiltonian
\begin{equation}
\begin{split}
\hat{H} = -\sum_{ij}{J_{ij} \hat{c}_{i\sigma}^\dag  \hat{c}_{j\sigma}} +\sum_{i}{U \hat{n}_{i\uparrow}\hat{n}_{i\downarrow}}+\sum_{ij\sigma\sigma^\prime}{V_{ij} \hat{n}_{i\sigma}\hat{n}_{j\sigma^\prime}}+\\+ \sum_{m} \omega_{m} \hat{b}_{m}^\dag \hat{b}_{m} + \sum_{ij\sigma} M_{il} \hat{n}_{i\sigma}\hat{x}_l\, 
\label{eq:hamiltonian}
\end{split}
\end{equation}
with $\hat{x}_l=\hat{a}_l+\hat{a}_l^\dagger$. Note also the presence of the two types of phonon operators $\hat{a}$, $\hat{b}$ left for brevity. We have neglected the impact of the terms that involve phonon-induced tunneling, as they are suppressed in the same way as the nearest-neighbor interaction terms due to small Wannier function overlap.

\begin{figure}
	\centering
	\begin{subfigure}[b]{0.45\textwidth}
		\includegraphics[width=0.9\textwidth]{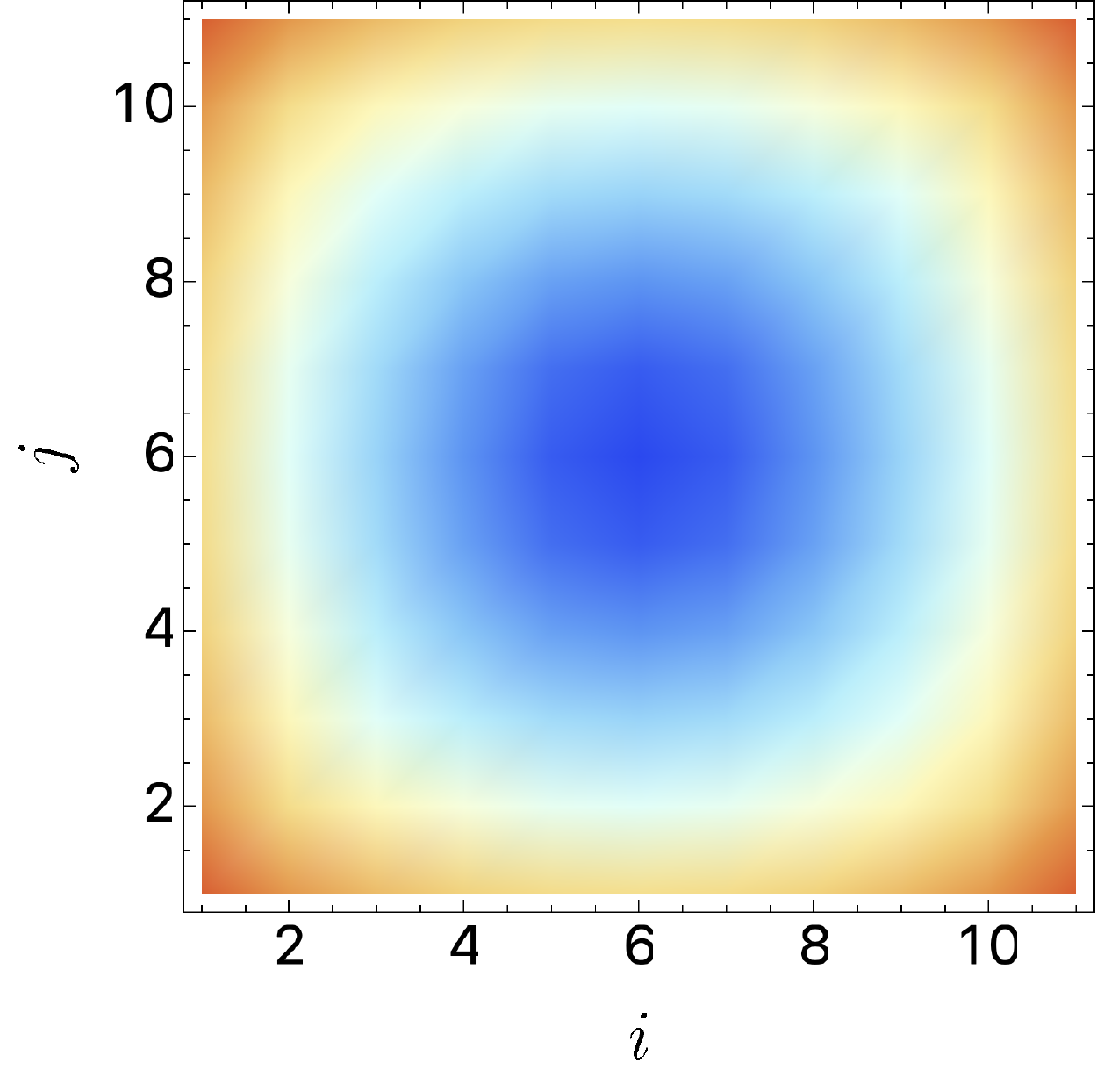}\\
		\includegraphics[width=0.6\textwidth]{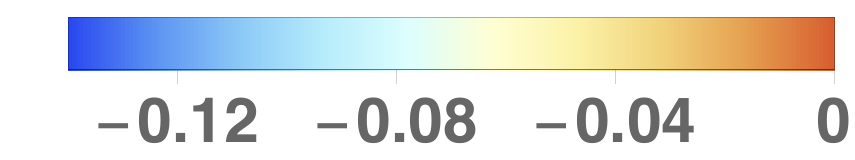}
	\end{subfigure}
	\begin{subfigure}[b]{0.45\textwidth}
		\includegraphics[width=0.9\textwidth]{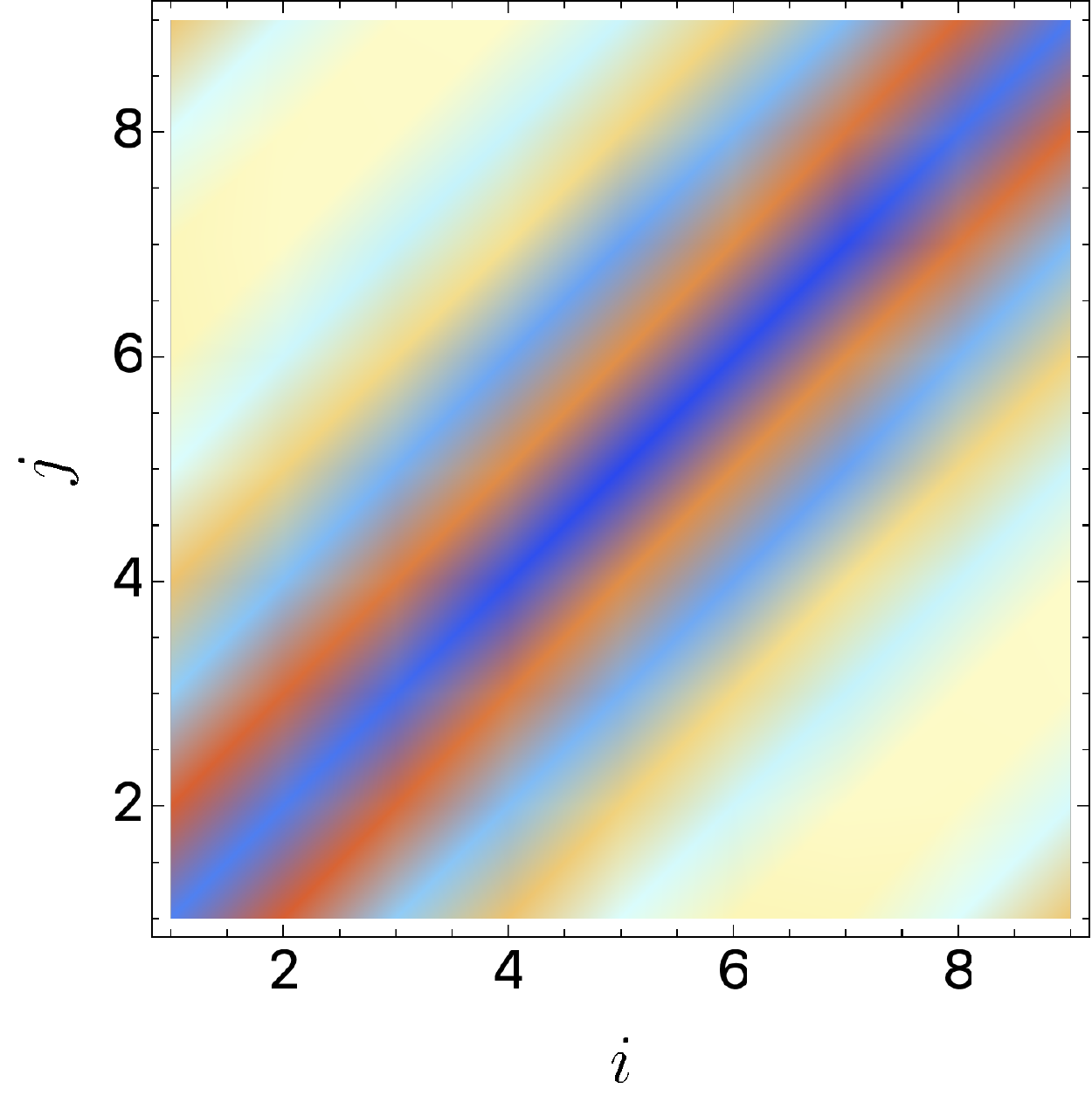}\\
		\includegraphics[width=0.7\textwidth]{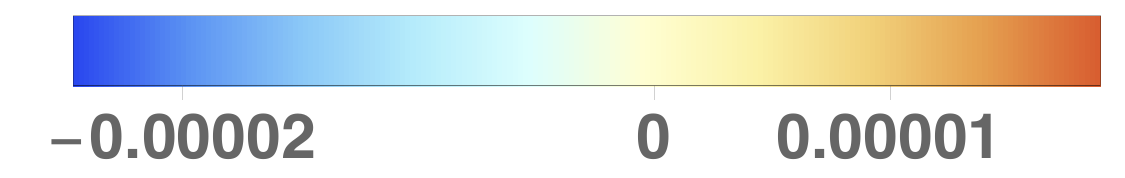}
	\end{subfigure}
	\caption{\label{fig:effint}The effective interaction between the atoms $W_{ij}/E^\star$ after the Lang-Firsov transformation for two different values of the ion-atom scattering length $a^e=0.1R^\star$ (a) and $a^e=0.008R^\star$ (b).}
\end{figure}

In order to get more insight into the physics of the ion chain, it is convenient to perform the generalized Lang-Firsov transformation $\overline{H}=e^S H e^{-S}$~\cite{Lang1963,Linden2004} defined by the generator
\beq
\hat{S}=i\sum_{i,j}{\lambda_{ij}\hat{n}_i(\hat{a}_j-\hat{a}_j^\dagger)}\, ,
\eeq
where $\lambda_{ij}$ are for now arbitrary complex numbers. The transformation, detailed in Appendix~\ref{appb}, introduces long-range phonon-mediated interactions and dresses the tunneling term with the lattice distortions. We choose the values of $\lambda$ parameters in such a way that the atom-phonon coupling term is canceled, which requires solving a system of linear equations due to nontrivial phonon mode structure of the chain. In contrast, for the case of purely local phonons and translational invariance one can eliminate the coupling term with a single $\lambda$ parameter that does not depend on the site index, namely $\lambda_{ij}=\delta_{ij}M_{ii}/\Omega$. After the transformation one obtains a new interaction term $W_{ij} n_i n_j$ with 
\beq
\label{eq:indint}
	W_{ij}={\left(\frac{U}{2}\delta_{ij}+\sum_{k}M_{ij}\lambda_{kj}+\frac{1}{4}\sum_{mkl}\omega_m\gamma_{km}^\star\gamma_{lm}\lambda_{ki}\lambda_{lj}\right)}\, ,
\eeq 
with $\gamma_{km}=u^m _k-b^m _k$. This is the long-ranged interaction mediated by the phonons, containing a direct coupling term, but also an additional one which arises due to the nontrivial mode structure of the crystal. Both terms can turn out to be important depending on the system parameters. Interestingly, the first term in Eq.~\eqref{eq:indint} contains the information about the coupling coefficient, while the second one involves the decomposition of the phonon modes, thus potentially leading to a difference between the bare coupling and the total induced interaction.
In Figure~\ref{fig:effint}, we show the effective interaction for the same parameters as in Fig.~\ref{fig:atph}. For small scattering length shown in panel (b), corresponding to strong interactions, we find that the effective term has the form of a decaying sinusoid, as found also in Ref.~\cite{Ortner2009} for atoms moving in a molecular crystal. Quite strikingly, in panel (a) the interactions have a completely different form, attracting each other weakly on long scales. This is due to the fact that the effective interaction is mediated mainly by the lowest phonon mode of the ionic lattice, as the other modes have much higher energies, and as a consequence can reflect the shape of a single collective mode. This is in contrast to the models employing local phonons, which always lead to the effective interactions of the shape shown in panel (b) of Fig.~\ref{fig:effint}. 

%
%
\section{Polaron properties}
%
The model described by Eq.~\eqref{eq:hamiltonian} has a considerably rich structure and represents a numerical challenge even in one dimension. Let us first briefly discuss the physics of a simpler model in which thel interactions, coupling to phonons and the phonon modes are strictly local. This way one arrives at the Hubbard-Holstein model given by the Hamiltonian
\beq
\begin{split}
	\hat{H}_{\rm H-H} = -J\sum_{ij\sigma}{\hat{c}_{i\sigma}^\dag  \hat{c}_{j\sigma}} +\sum_{i}{U \hat{n}_{i\uparrow}\hat{n}_{i\downarrow}}+\\ + \omega_0 \sum_{i}  \hat{a}_{i}^\dag \hat{a}_{i} +M_0 \sum_{i\sigma} \hat{n}_{i\sigma}\hat{x}_i\, .
	\label{eq:hh}
\end{split}
\eeq
This Hamiltonian can be viewed as the simplest possible extension of the Hubbard model taking into account the coupling to phonons and has been widely studied in the literature~\cite{Sil1996,Bonvca1999,Takada2003,Linden2004,Clay2005,Goodvin,Tezuka2007,Fehske2008,Wang2019,Reinhard2019}. Its main feature is the competition between the coupling term and the interaction, which drive the system towards two different phases, the Mott insulator and charge density wave. Furthermore, an additional phase can emerge at the interface of the two insulators. Surprisingly, this intervening phase has been shown to be conducting~\cite{Takada2003}. Extending the model~\eqref{eq:hh} e.g. by including long-range interactions can lead to even richer physics, e.g. induce pairing between the polarons and potentially turn the metallic phase into a superconducting one~\cite{Campbell2014}. Naturally, the most interesting and computationally challenging problem emerges when all the terms in the Hamiltonian compete with each other.

As shown in Sec.~\ref{sec:atph}, the atom-ion platform is suitable for realizing a wide class of extended Hubbard-Holstein models, as both the magnitude and range of the atom-phonon coupling can be tuned in experiment. Multiple available control knobs make the system versatile. We thus expect that it can serve as an alternative to numerical methods, which have proven to be extremely challenging, for studying the phase diagram and the out-of-equilibrium dynamics of such models by means of analogue quantum simulation. Let us consider the example of the standard Hubbard-Holstein model as given by Eq.~\eqref{eq:hh}. Here one could e.g. set the phonon frequency $ \omega_0$, tunneling coefficient $J$ and the phonon coupling $M_0$ as constant and manipulate the atomic interaction $U$ by means of a Feshbach resonance. Going from strong to weak repulsion, the system should go from the Mott insulator to a charge density wave phase via the intermediate metallic phase. Readily developed detection methods such as absorption imaging and time-of-flight would provide access to the quantities needed to distinguish and characterize these phases, such as the number fluctuations on a single lattice site, two-point correlations and the momentum distribution.

Let us now briefly discuss the emerging physics on the level of one or two fermions, neglecting the phonon dynamics. After the Lang-Firsov transformation, the atoms become dressed with the lattice phonons. These new quasiparticles, called polarons, are now characterized by a new dispersion relation and modified interactions. Properties and dynamics of polarons and their possible bound states (bipolarons) can be studied more easily than the full many-body system, and provide useful information. For instance, existence of mobile bipolarons is a precursor for pairing. 
This is especially relevant in the strong coupling regime in which the coupling to phonons dominates over the tunneling, where the systemcan have substantially different properties from the Hubbard model. 

In order to demonstrate the impact of phonon-induced interactions, let us discuss the effective mass of the bipolaron, which is a bound state of two dressed fermions. For a lattice model with purely local interactions, the only possibility for the bound state to exist is in the spin singlet state when the effective onsite interaction taking the phonon contribution into account is attractive. In contrast, adding a finite range interaction leads to two additional states where the fermions are bound in neighboring sites, regardless of their spin state. This intuitively provides an additional pairing mechanism, which can enhance the conductivity~\cite{Campbell2014}. Furthermore, the effective mass of the singlet localized bipolaron is also strongly affected by the phonons, especially if the induced interaction is nonlocal~\cite{Bonvca2001}. The impact of nearest neighbor interaction term on the singlet bipolaron mass is studied in Appendix~\ref{appc}. In brief, even if the bound state is localized on a single lattice site, the second order correction to its energy comes from two virtual hopping events to the adjacent site and back. The intermediate state with two populated sites is then modified by the interaction term. As a result, the bipolaron becomes exponentially lighter by a large factor $\propto \exp{(\epsilon_P/\omega_0)}$, where $\epsilon_P\propto M_0$ denotes the polaron energy.
Our calculations indicate that observation of such effects in a hybrid ion-atom system can be possible. Indeed, the typical kinetic energy $J$ can take values of the order of $0.1\, E^\star$ as shown in Fig.~\ref{fig:setup}(b), the phonon frequency $\omega_0$ as well as the bare atomic interaction $U$ can be manipulated via the external potential and Feshbach resonances, and the induced terms can reach $0.1\, E^\star$ as it can be seen in Fig.~\ref{fig:effint}, leading to high tunability of the ratio between the local and nonlocal interactions. In principle, even small ion chains with $N\geq 5$ can be sufficient for the observation of bipolaron states, as they would be strongly localized.

%
\section{Conclusions}
%
In this work, we outlined a possibility for quantum simulation of solid state models in which the coupling to phonons competes with strong interactions between fermions. Our proposal relies on preparing a hybrid ion-atom system with a chain of ions acting as a lattice for the atoms. While ion chains are now routinely prepared in laboratories, so far ultracold ion-atom hybrid systems have only been created using a single ion~\cite{TomzaRMP}. From the atomic side, state-of-the-art atomic setups provide the possibility to control the number of atoms to a high degree~\cite{Wenz2013}. In order to gain access to the phonon spectrum, one needs an exceptional level of control over the system, but in principle the microscopic parameters of the model can be tuned via manipulating the trap geometry and the interparticle interactions. The resulting parameters such as induced interactions can reach a fraction of the characteristic energy $E^\star$, meaning that observation of interesting quantum phases will require cooling the system to nanokelvin temperatures. Recent experimental advancements~\cite{Kleinbach2018,Feldker2019,Karpa2020} indicate that this can be achieved in the near future.

We have considered a quasi-one-dimensional system consisting of a stable linear ion chain. Even in this simplistic case, the study of the phase diagram and the role of induced nonlocal interactions would be relevant and complementary to the numerical calculations. It would also be interesting to investigate the quantum phase transition to a zig-zag geometry, which would lead to realization of a two-leg ladder lattice. The particularly interesting region is the phase transition point itself, where quantum fluctuations of the ions are strong~\cite{ShimshoniPRL2011}. Coupling to the atomic gas would presumably affect the excitation spectrum in this region and could affect the critical properties. This could be controlled in experiment by tuning the transverse separation between the linear ion crystal and the fermionic gas or by laser-controlled long-ranged atom-ion interactions~\cite{Secker2016}. Further research directions include out-of-equilibrium scenarios such as studying the dynamics of (bi)polarons after quantum quenches and preparation of nontrivial phonon states, e.g. by squeezing them through external driving, possibly making the connection to light-induced superconductivity~\cite{Mitrano2016}.

\section*{Acknowledgments}

We would like to thank Yao Wang, Eugene Demler, and Rene Gerritsma for insightful discussions and ITAMP for hospitality and support via the Visitor Program. This work was supported by the DFG SPP 1929 (GiRyd), the Polish National Agency for Academic Exchange (NAWA) via the Polish Returns 2019 programme and the Cluster of Excellence `CUI: Advanced Imaging of Matter' of the Deutsche Forschungsgemeinschaft (DFG) - EXC 2056 - project ID 390715994.

\appendix
\section{Hamiltonian derivation details}
\label{appa}
\subsection{Phonon modes}

A convenient approach to calculate the phonon spectrum for an arbitrary array of ions has been provided by Bissbort {\it et al}~\cite{Bissbort2016}. The procedure begins by finding the classical equilibrium positions of the ions, and expansion of the potential energy around the equilibrium up to the second order $V_{ij}=\frac{\partial^2 V}{\partial \left(\delta R_i\right)\partial \left(\delta R_j\right)}$ (calculated at equilibrium). The local harmonic oscillator frequency for the $j$th ion can then be defined as
\beq
\Omega_j=\sqrt{\frac{V_{jj}}{M}},
\eeq
where $M$ denotes the mass of each ion (assuming identical ions for simplicity). In the next step one introduces local ladder operators corresponding to the local oscillators (in the Appendices we drop the hats usually denoting the operators)
\beq
\beta_l=\sqrt{\frac{M \Omega_l}{2}}\left(\delta R_l+\frac{i}{M \Omega_l}P_l\right)\, .
\eeq
Rewriting the Hamiltonian in terms of the local operators results in a quadratic theory $H_{\rm ph}=E_0+\frac{1}{2} \left(\begin{array}{c}\beta \\ \beta^\dagger\end{array}\right)^\dagger \mathcal{H}_{\rm ph} \left(\begin{array}{c}\beta \\ \beta^\dagger\end{array}\right)$, which can be diagonalized using generalized Bogoliubov transformation on symplectic space. The right eigenvectors have the form $\mathbf{x}^m=\left(\begin{array}{c}\mathbf{u}^m \\ -\mathbf{v}^m\end{array}\right)$. The diagonalized Hamiltonian describes collective phonon modes
\beq
H_{\rm ph}=\sum_m{\hbar\omega_m b_m^\dagger b_m}
\eeq
with $\omega_m$ being the energy of a collective mode $m$, and $b_m$, $b_m^\dagger$ denoting the phonon creation and annihilation operators which fulfill the bosonic commutation relations. By inverting the transformation, they can be connected to the local ion displacement operators $\delta R_j$ via
\beq
\label{deltaR}
\delta R_j=\sqrt{\frac{\hbar}{2M \Omega_j}}\sum_m{\left[(u_j ^m-v_j ^m)b_m+(u_j ^m-v_j ^m)^\ast b_m ^\dagger\right]}\, .
\eeq 

\subsection{Atomic band structure}
The band structure for an atom moving in the periodic potential provided by the ion chain has been calculated in Ref.~\cite{Negretti2014}. It is sufficient to assume static ions at their equilibrium positions, while the coupling to phonons will be described in the next section. The one-dimensional ion-atom interaction can be written using the following general pseudopotential
\beq
V_{\rm a-i}(x)=g^e\delta(x)+g^o\delta^\prime(x)\partial_\pm
\eeq
with $g^e=-\hbar^2/\mu a^e$ and  $g^o=-\hbar^2 a^o/\mu$ describing the even and odd part of the interaction and $a^e$, $a^o$ are the one-dimensional energy-dependent scattering lengths. 

The lattice Fourier transform of the potential reads 
\beq
V_{\rm a-i}(q)=(g^e+g^o q^2)/d\, .
\eeq
In a periodic chain with ions separated by distance $d$, the band dispersion is given by
\beq
\cos kd=\frac{a^e+a^o}{a^e-a^o}\cos qd +\frac{q^2 a^e a^o -1}{(a^e-a^o)q}\sin qd\, .
\eeq

The quasimomentum vector in a finite lattice is quantized $k=2\pi n/(N d)$ with integer $n=0,\, \pm 1,\, \pm 2\,\dots\, \pm(N-1)/2$. The normalized Bloch functions for periodic boundary conditions can be found analytically~\cite{Negretti2014}. The Wannier functions are defined as lattice Fourier transform of the Bloch states. In order to maximize the localization of the Wannier states, we use Kohn's prescription~\cite{Kohn1959}, namely we multiply the Bloch functions by a constant phase factor computed such that $\Im\psi_k(0)=0$.

\section{Lang-Firsov transformation}
\label{appb}

The full Hamiltonian taking into account the ions, atoms and the coupling to phonons has the generic form
\beq
\begin{split}
H=-J\sum_{\langle i,j\rangle\sigma}c_{i\sigma}^\dagger c_{j\sigma}+U\sum_j{n_{j\uparrow}n_{j\downarrow}}+\\+\sum_m{\omega_m b_m^\dagger b_m}+\sum_{jl}{M_{jl}n_j\left(a_l+a_l^\dagger\right)},
\end{split}
\eeq
where the atom-phonon coupling term is written using local oscillators $a_j$, in contrast to the collective phonon modes $b_j$ and we have omitted the phonon-induced tunneling term. We now  perform the generalized Lang-Firsov transformation, which enables clear identification of different parameter regimes as well as further numerical treatment. The conventional version of the transformation, employed for local atom-phonon coupling, is generated by the operator $S=i\sum_j \lambda_j n_j (a_j-a_j^\dagger)$ with $\lambda$ being an adjustable parameter. This form is sufficient, e.g., to eliminate the atom-phonon coupling term from the transformed Hamiltonian. However, in our case we need the generalized version utilizing $S=i\sum_{jl}\lambda_{jl}n_j (a_l-a_l^\dagger)$. The transformed operators are easy to write down, e.g. $\tilde{c}_j=c_j \exp{i\sum_l}\lambda_jl(a_l-a_l^\dagger)$,  $\tilde{a_l}=a_l+\frac{1}{2}\sum_j{\lambda_{jl}n_j}$, but one has also to include the decomposition of the phonon modes into the local displacements $b_m=\sum\gamma_{im} a_i$. The transformed Hamiltonian takes the form
\begin{widetext}
\beq
\begin{split}
	\tilde{H}=-J\sum_{\langle i,j\rangle\sigma}{c_{i\sigma}^\dagger c_{j\sigma}\exp{\left(i\sum_l(\lambda_{il}^\ast-\lambda_{jl})(a_l-a_l^\dagger)\right)}}+\sum_m{\omega_m b_m^\dagger b_m}-\frac{U}{2}\sum_i n_i+ \\
	+ \sum_{jl}{\left(M_{jl}+\sum_{km}\frac{1}{2}\omega_m\gamma_{lm}^\ast \gamma_{km}\lambda_{kj}\right)n_j(a_l+a_l^\dagger)} + \sum_{ij}n_i n_j{\left(\frac{U}{2}\delta_{ij}+\sum_{k}M_{ij}\lambda_{kj}+\frac{1}{4}\sum_{mkl}\omega_m\gamma_{km}^\ast \gamma_{lm}\lambda_{ki}\lambda_{lj}\right)}\, .
\end{split}
\eeq
\end{widetext}
The characteristic features of the model in this reference frame are: the modification of the tunneling term (dressing of atoms by the lattice distortion), the emergence of long-range interactions due to the nonlocal atom-phonon coupling term (both directly and indirectly from the collective character of the phonon modes), and a modified atom-phonon coupling term.

It is now possible to choose the $\lambda_{ij}$ coefficients in such a way that the atom-phonon coupling term is completely eliminated, leaving behind a potentially long-range effective interaction between the atoms.  However, it is also possible to optimize the values of $\lambda$ in a different way, e.g. to minimize the energy of some variational wavefunction.

In order to get more insight here, it is useful to consider the simplified version of the model with translational invariance, local phonons with frequency $\omega_0$ and coupling strength $M_0$ and no long-range coupling terms~\cite{Linden2004}. In this case one has $\lambda=M_0/\omega_0$ and the transformed Hamiltonian reads
\beq
\begin{split}
\tilde{H}=-J\sum_{\langle i,j\rangle\sigma}{c_{i\sigma}^\dagger c_{j\sigma}\exp{\left(i\lambda(a_i-a_i^\dagger-a_j+a_j^\dagger)\right)}}+ \\ +\omega_0\sum_i{a_i^\dagger a_i}-\frac{U}{2}\sum_i n_i +\sum_{ij}{n_i n_j\delta_{ij}\left(\frac{U}{2}-\frac{M_0^2}{\omega_0}\right)}
\end{split}
\eeq
with the onsite interaction shifted by $E_P=M_0 ^2/\omega_0$. Also in the case of the full model the induced interactions are expected to be of the order of $M_0^2/\omega_0$ with the lowest phonon frequency inducing the strongest interaction.

Let us discuss several regimes which can simplify the situation. Firstly, the adiabaticity parameter $\epsilon=\omega_0/J$ can be defined. In the antiadiabatic regime $\epsilon\gg 1$ the phonons are ``fast'' compared to the atoms and they can be integrated out leading to an extended Bose-Hubbard model with renormalized hopping and long-range interactions. In general, one can assume the phonons to decouple from the atoms and use a simple wavefunction such as the vacuum state, a thermal state or a coherent state for the phononic part. This approximation gets worse as the phonon energy scales become comparable to the atomic part of the Hamiltonian. It is also worth noting that in the case of a discrete phonon spectrum the lowest modes are decisive for the shape and strength of the interaction. In the antiadiabatic scenario, if $M_0^2/\omega_0 \ll 1$, the corrections induced by the phonons are negligible. In our setup one can tune the lowest mode frequency in a wide range, allowing to switch between different regimes. Furthermore, in a short chain already the second lowest mode is separated in energy scale from the other quantities, making the lowest mode the only one coupled to the atomic dynamics. This can be seen in the spatial profile of the induced interaction, which is reminiscent of the ionic displacement amplitudes of the lowest mode as demonstrated in Fig.~\ref{fig:effint}.

\section{Strong coupling expansion}
\label{appc}

Here, we briefly discuss the calculation of the effective mass in the strong coupling regime based on Refs.~\cite{Bonvca1999,Bonvca2001}. After performing the Lang-Firsov transformation  as in the previous section, we consider the case of a single atom or two atoms in the chain. For simplicity we neglect finite size effects and work with translationally invariant model described by local phonons with frequency $\omega_0$ and coupling strength $M_0=-\omega_0 g^2$ with dimensionless $g$ parameter. Here we also take $\hbar=1$ and the lattice spacing $d=1$ to simplify the notation. We assume that the wave function separates into an atomic and phononic part, and take the phonon state to be the vacuum. This leads to a general model with the interaction term described by
\beq
\label{eq:int}
H_0=U\sum_i{n_{i,\uparrow}n_{i,\downarrow}}-\omega_0 g^2\sum_{ijl}f_l(i)f_l(j)n_i n_j\, ,
\eeq
and the kinetic part by
\beq
\label{eq:kin}
T=-Je^{-\tilde{g}^2}\sum_{j\sigma}{c_{j+1,\sigma}^\dagger c_{j,\sigma}e^{-g\sum_l{(f_l(j+1)-f_l(j))(a_l^\dagger-a_l)}}}+h.c.
\eeq
Here $f_l(i)$ describes the coupling of the atom at site $i$ to the oscillator at site $l$ and can be nonlocal. For the Holstein-Hubbard model we have $f_l(i)=\delta_{li}$. In our system the coupling to the next nearest neighbours is non-negligible. We will consider a simplified model in which 
\beq
f_l(i)=\delta_{li}+\kappa(\delta_{l+1,i}+\delta_{l-1,i})
\eeq
with $0<\kappa<1$ being the model parameter. The band narrowing factor $e^{-\tilde{g}^2}$ in Eq.~\eqref{eq:kin} is given by
\beq
\tilde{g}^2=g^2\sum_{l}{\left[f_l(0)^2-f_l(0)f_l(1)\right]}
\eeq
and results in $g^2$ for the Holstein case, but for our model is equal to $g^2(2\kappa^2-2\kappa+1)$. The polaron shift (the correction to the onsite interaction from the last term in Eq.~\eqref{eq:int}) $\epsilon_P=\omega_0 g^2$ for the Holstein model, while in the present model we have
\beq
\epsilon_P=\omega_0 g^2(1+4\kappa+4\kappa^2)\, .
\eeq
This gives the ratio $\omega_0 \tilde{g}^2/\epsilon_P=\frac{1-2\kappa+2\kappa^2}{1+4\kappa+4\kappa^2}$, in contrast to $1$ for the Holstein case.

In the strong coupling limit $g\gg 1$, the kinetic term can be regarded as a perturbation. The energy can then be calculated using perturbation theory. For a single atom in the first order we have $E_p=-\epsilon_P-2Je^{-\tilde{g}^2}\cos k$. The second order correction to this expression reads
\beq
E_p^{(2)}=-J^2 e^{-2\tilde{g}^2}\sum_n{\frac{\left((1-\kappa)g\right)^{2n}}{n! n\omega_0}(2+2\cos k)^2}\, .
\eeq
One can now calculate the effective mass of the polaron $m_p=\lim_{k\to 0}\left(\frac{\partial^2 E(k)}{\partial k^2}\right)^{-1}$, which results in
\beq
\begin{split}
m_p^{-1}=2Je^{-\tilde{g}^2}\left[1+\frac{8 J e^{-\tilde{g}^2}}{\omega_0}\left( \Gamma[0, g^2(1-\kappa)^2] + \right.\right.\\ \left. \phantom{\frac{1}{1}}\left. -\log[g^2(1-\kappa)^2]-\gamma\right)\right]
\end{split}
\eeq
with $\gamma$ denoting Euler's gamma constant. At this level the model does not really differ from the Holstein polaron mass apart from the modification of the coupling strengths $g$ and $\tilde{g}$. In the limit $g\to\infty$ the mass approaches the limit $e^{\tilde{g}^2}/2J\left(1-8J/\omega_0 g^2(1-\kappa)^2\right)$.

The case of the bipolaron is more involved. We expect that two polarons can create a bound state if there is some effective attraction in the system. For the Holstein model this arises when the effective onsite interaction $\tilde{U}=U-2\omega_0 g^2<0$, and a localized bipolaron is the ground state. Close to the $\tilde{U}=0$ point, a state extending over two lattice sites localized due to the tunneling exchange can also exist. 
Nonlocal atom-phonon coupling changes the situation significantly, as due to the induced interaction polarons can bind across the adjacent lattice sites. Furthermore, a bound state can also exist even if the atoms are in a triplet spin state so that they cannot occupy the same site, which is not possible in the Holstein model. The important difference is that due to the coupling to adjacent phonons, the mass of the bipolaron is much lower than in the Holstein case. To see this, let us consider the simplest case of the singlet localized bipolaron, for which up to the first order we have $E_b=U-2\epsilon_P$. The second order correction is given as
\beq
E_b^{(2)}=4J^2e^{-2\tilde{g}^2}\sum_{n,m}{\frac{(g(1-\kappa))^{2(n+m)}(1+(-1)^{n+m}\cos k)^2}{n!m!(n+m)(\omega_0-U+2\epsilon_P-\omega_0 g^2 \kappa)}}\, ,
\eeq
where the denominator is modified by the interaction of polarons in adjacent lattice sites. The leading term in the singlet bipolaron mass in the $g\to\infty$ limit reads
\beq
m_b\stackrel{g\to\infty}{\longrightarrow}e^{2\tilde{g}^2+2g^2(1-\kappa)^2} \frac{g^2 \omega_0+0(1-\kappa)^2}{4 J^2 u (u-\omega_0)}\, ,
\eeq
where we introduced $u=U-2\epsilon_P+\omega_0 g^2\kappa$ to shorten the notation. It is interesting to note that while for the Holstein bipolaron its mass is proportional to $\exp{4g^2}$, which is equivalent to $\exp{4\epsilon_P/\omega_0}$, here we have $m_b\propto\exp{4\xi\epsilon_P/\omega_0}$ with $\xi=(1-2\kappa+3\kappa^2/2)/(1+2\kappa)^2$ being a correction coefficient. The bipolaron in our model can thus be much lighter, i.e. by a factor $\exp{2\epsilon_P/\omega_0}$ already at $\kappa\approx 0.12$ for which $\xi=0.5$. 

\bibliography{liter}

\end{document}